\documentclass[prl,twocolumn]{revtex4}

\usepackage{graphicx}% Include figure files
\usepackage{bm,color}% bold math

\usepackage{graphicx}
\usepackage{amsmath}
\usepackage{bm,color}
\usepackage{multirow}

\newcommand{\be}{\begin{eqnarray}}
\newcommand{\ee}{\end{eqnarray}}

\usepackage{ulem}
\usepackage{xcolor}

\begin{document}

\title{Flat Band Quantum Scar}
\date{\today}

\author{Yoshihito Kuno}
\author{Tomonari Mizoguchi}
\author{Yasuhiro Hatsugai}
\affiliation{Department of Physics, University of Tsukuba, Tsukuba, Ibaraki 305-8571, Japan}

\begin{abstract}
We show that a quantum scar state, an atypical eigenstate breaking eigenstate thermalization hypothesis embedded in a many-body energy spectrum, can be constructed in flat band systems. 
The key idea of our construction is to make use of orthogonal compact localized states.
We concretely discuss our construction scheme, taking a saw-tooth flat lattice system as an example, and numerically demonstrate the presence of a quantum scar state. Examples of higher-dimensional systems are also addressed. Our construction method of quantum scar has broad applications to various flat band systems.
\end{abstract}

%\pacs{67.85.Hj, 75.10.-b, 03.75.Nt}

\maketitle
%%%%%%%%%%%%%%%%%%%%%%%%%%%%%%
%%%%%%%%%%%%%%%%%%%%%%%%%%%%%%
{\it Introduction.---}
Violation of eigenstate thermalization hypothesis (ETH) \cite{Deutsh,Srednicki,Rigol,Rigol2} now attracts great interest, being called weak ETH \cite{Alessio,Gogolin}.
The violation leads to area-law entanglement entropy (EE) for some specific eigenstates, 
while almost all eigenstates exhibit thermalization and volume-law EE. 
Recently, a Rydberg coldatom quantum simulator heuristically accessed specific eigenstates inducing the violation of the ETH through a quench dynamics \cite{Bernien}. A suitable nonentangled state (Charge density wave state) has not thermalized during a long-time evolution and a recurrence occurs, indicating that the initial information is not lost. 
Immediately, the theoretical model describing the Rydberg coldatom quantum simulator, namely the PXP model, has been studied in detail and the study clarified that the series of the non-thermalized eigenstates with the are--law EE exists, but the system is nonintegrable as a whole (there are no extensive numbers of the local conserved quantities) and a quench dynamics exhibits a coherence, corresponding to very--slow thermalization and linear (slow) growth of EE \cite{Turner,Turner2,Choi,Ho,Lin,Khemani}. Such specific eigenstates not possessing the typical properties from the ETH are buried in most thermal eigenstates \cite{Shiraishi}. They are now called quantum scars (QSs), whose the single-particle counterpart has been reported four decades ago \cite{Heller}.

QSs can appear not only for the PXP model but also for broad condensed--matter models.
In fact, exact eigenstates with long-range orders away from the ground states were known in the literature, such as an $\eta$-pairing state in a Hubbard model~\cite{Yang1989,Zhang1990},
and indeed revisited as a candidate of QS~\cite{Vafek2017}. 
Recently, a general construction has been proposed \cite{Shiraishi} before the first Rydberg experiment \cite{Bernien}.
Since then, an increasing number of examples have been reported, including the AKLT model \cite{Moudgalya1,Moudgalya2}, some $S=1$ spin XY models \cite{Schecter,Chattopadhyay}, 
 topological models \cite{Ok,Srivatsa}, frustrated spin systems \cite{McClarty}, a quantum dimer model \cite{Wildeboar},
and disordered systems \cite{Shibata,Hart}.

In this Letter, we propose a simple general construction of QS states in flat band systems, 
where the model has a spatially compact localized state (CLS) \cite{Leykam,Santos2020,Sathe} as eigenstates on the flat band, 
which does not spatially overlap, and thus are orthogonal with each other. 
This construction of the QS has a broad application to various flat band systems.
By making use of ortho-normalized CLSs, we can construct a low entangled many--body state violating the ETH, which remains to be an exact eigenstate even in the presence of conventional density-density interaction; other states obey the ETH as the models themselves are generally nonintegrable. 
In the following, we first present a generic argument mentioned above, and then present a concrete example of 
the saw-tooth-lattice model. 
We numerically demonstrate the realization of the QS state by showing the level statistics and time-evolution of the entanglement.  
In addition, some examples of a two-dimensional model are presented.

%%%%%%%%%%%%%%%%%%%%%%%%%%%%%%%%%%%%
%Fig
%\widetext
\begin{figure}[b]
\centering
%\begin{center}
%\centering  
\includegraphics[width = 0.98\linewidth]{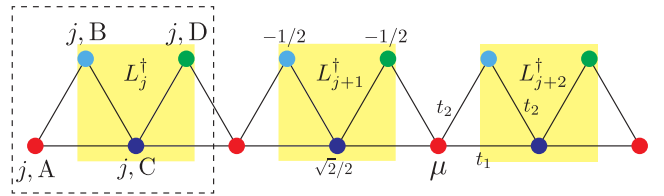}
%\includegraphics[width=7.5cm]{Fig1.eps}
%\end{center} 
% \hspace{10truemm}
\caption{Schematic picture of saw-tooth lattice. The yellow shade object is a CLS. 
The CLS has no spatially overlap each other due to the presence of a finite $\mu$. The minimum distance between the neighboring CLS is $d_{m}=2$.}
\label{Fig1}
\end{figure}
%%%%%%%%%%%%%%%%%%%%%%%%%%%%%%%%%%%%

{\it General construction.---}
We propose a general construction for a unique QS.
We start with considering the following general flat band model 
with interactions for spinless fermions \cite{gene_ham}. 
Let us consider a lattice with $N_{\rm t}$ sites composed of a periodic array of $N_{\rm L}$ unit cells;
there are $N_{\rm t}/N_{\rm L}$ sublattices per unit cell.
On this lattice, we consider the Hamiltonian:
\begin{eqnarray}
H_{\rm tot}&=&H_{\rm 0} + H_{\rm int},\\
H_{\rm 0} &=& \sum_{ij}f^{\dagger}_{i}h_{ij}f_{j},\:\: 
H_{\rm int} = \sum_{|i-j|\leq \ell}V_{ij}n_{i}n_{j},
\end{eqnarray}
Here, $H_{\rm 0}$ stands for the single-particle Hamiltonian hosting a flat band 
with $f^{(\dagger)}_{i}$ being an annihilation (creation) fermion operator in real space, and $H_{\rm int}$ is a two--body finite-range interaction with the maximum range $\ell$, $V_{ij}$ is an interaction strength, and $n_{j}=f^{\dagger}_{j}f_j$. 
It is to be emphasized that no elaboration of $H_{\rm int}$ is needed to obtain the QS state.
Rather, the key feature to obtain the QS state is encoded in $H_{\rm 0}$, that is, we assume that the flat band eigenstates are spanned by {\it ortho-normalized} localized states which have a compact support and do not spatially overlap each other;
such states are referred to as the CLSs (see yellow shades in Fig.~\ref{Fig1} as an example). 
It is noteworthy that not all the flat band models satisfy this assumption. 
In fact, in some models, such a set of localized states either overlap each other or do not have a compact support~\cite{Bergman2008,Huber}.

Let $L_j$ be an annihilation operator of the CLS at the unit cell $j \in[0,N_{\rm L}-1]$,
and $d_m$ be the minimum of the distance between the sites involved in the supports of neighboring $L_j$'s,
assumed to satisfy $d_m > 1$. 
Then, for the many-body system with particle number being fixed to $N_{\rm L}$,
we consider the following state:
\begin{eqnarray}
|\Psi_{\rm L} \rangle =\prod^{N_{\rm L}-1}_{j=0} L^{\dagger}_{j}|0\rangle.  \label{eq:scarstate}
\end{eqnarray} 
This state is created by occupying the entire eigenstates on the flat band.
Notably, for a finite-ranged $H_{\rm int}$, we can keep the state $|\Psi_{\rm L}\rangle$ an exact eigenstate due to the isolation of the CLS state. 
Namely, if the interaction $H_{\rm int}$ is a two--body interaction under the condition $\ell< d_{m}$, 
the state $|\Psi_{\rm L}\rangle$ is still eigenstate for the interacting system, because $H_{\rm int}|\Psi_{\rm L}\rangle =0$, 
while $H_{\rm int}$ convert the total system into nonintegrable.
In the literature, exact many-body eigenstates with vanishing interaction energy were considered in the context of flat band ferromagnetism~\cite{Mielke1991,Tasaki1992,Tasaki1998}, and later in the Wigner crystal~\cite{Wu}.
As a result, $|\Psi_{\rm L}\rangle$ is a unique QS, which originates from the flat band nature, and we expects that $|\Psi_{\rm L}\rangle$ satisfies area--law scaling for the EE and is embedded in most thermal eigenstates in a system with a translational invariance. 
Further, the total system does not have extensive numbers of conserved quantities, so the total system is nonintegrable.

It should be noted that our construction has a relation to the recent general construction method 
by Shiraishi and Mori \cite{Shiraishi}, but our construction can be regarded as a slightly relaxed version; see Supplemental Material~\cite{Supp}.

{\it Example: Saw--tooth lattice.---}
By applying the above general argument, we show a concrete construction of the QS from CLSs, which is analytically very simple.
We start with the following model defined on the saw--tooth lattice (Fig.~\ref{Fig1}),
$H_0=\sum^{L-1}_{j=0} \biggl[t_1 f^{\dagger}_{j,\mathrm{A}} f_{j,\mathrm{C}}+t_2f^{\dagger}_{j,\mathrm{A}}f_{j,\mathrm{B}}
+t_2f^{\dagger}_{j,\mathrm{B}}f_{j,\mathrm{C}}
+t_1 f^{\dagger}_{j,\mathrm{C}} f_{j+1,\mathrm{A}}+t_2f^{\dagger}_{j,\mathrm{C}}f_{j,\mathrm{D}}
+t_2f^{\dagger}_{j,\mathrm{D}}f_{j+1,\mathrm{A}}
+\mbox{h.c.}\biggl] + \sum^{L-1}_{j=0}\mu f^{\dagger}_{j,\mathrm{A}}f_{j,\mathrm{A}}$,
and $H_{\rm int}$ is a nearest-neighbor interaction, 
given later (See Eq.~(\ref{int_ST})). 
Note that the finite on-site potential, $\mu \neq 0$, leads to the increase of the sublattice degrees of freedom from two to four. 
Further, for $t_2=\sqrt{2}t_1$, one of the bands out of four becomes a flat band; see Supplemental Material~\cite{Supp}
for the single-particle spectrum. 
The existence of the flat band can be inferred from the molecular-orbital (MO) representation~\cite{Supp},
which was developed to describe generic flat band models in the prior works~\cite{Maruyama, Mizoguchi}. 
We can also straightforwardly find {\it ortho-normalized} CLSs as 
\begin{eqnarray} 
L^\dagger_j = \frac{1}{2}\biggl[\sqrt{2}f^{\dagger}_{j,\mathrm{C}}-f^{\dagger}_{j,\mathrm{B}}-f^{\dagger}_{j,\mathrm{D}}\biggr], \:\: \{ L_{j},L^\dagger_{j'}\}=\delta_{jj'},
\end{eqnarray}
where $L_j$'s do not overlap each other, 
and satisfy $[L^\dagger_j ,H_0] = -2t_1 L^\dagger_j$.

Having these CLSs at hand, we now construct a many--body state, 
which turns into a QS when switching on interactions. 
We consider $1/4$ filling, 
then can construct the many-body state that occupies the entire states of the flat band,
given in the form of Eq.~(\ref{eq:scarstate}).
Clearly, this state is an exact eigenstate since 
$H_{0}|\Psi_{\rm L}\rangle = -2t_1 L |\Psi_{\rm L}\rangle$ and $H_{\rm int} |\Psi_{\rm L}\rangle = 0$.
This fact is independent of the value of $\mu$ and the profile of $V_{ij}$ ($|i-j|=1)$.
Further, since $H_{\rm int}$ is nearest-neighbor interaction and $d_m$ for $L_j$'s is equal to two,
$|\Psi_{\rm L}\rangle$ is the eigenstate of $H_{\rm int}$ with zero eigenvalue, thus, $|\Psi_{\rm L}\rangle$ remains as a many--body eigenstate of $H_{\rm tot}$.
The Wigner-solid-like particle distribution of $|\Psi_{\rm L}\rangle$ indicates that the state exhibits area--law EE.
This state is out of friends because the other many--body eigenstate are thermal delocalized, which are expected to obey the ETH and exhibit volume-law EE.
In the following, we numerically demonstrate that $|\Psi_{\rm L}\rangle$ is the QS embedded in this model.

{\it Numerical demonstration.---}
Let us numerically verify that $|\Psi_{\rm L}\rangle$ is the QS state~\cite{Quspin}. 
For concreteness, we set the profile of $H_{\rm int}$ as 
\begin{eqnarray}
H_{\rm int}&=&\sum_{j}V_0 \left(n^{\mathrm{A}}_j n^{\mathrm{B}}_{j}-n^{\mathrm{B}}_j n^{\mathrm{C}}_{j} 
+ n^{\mathrm{C}}_j n^{\mathrm{D}}_{j} - n^{\mathrm{D}}_j n^{\mathrm{A}}_{j+1} 
 \right),
 \label{int_ST}
\end{eqnarray}
where $n^{\alpha}_j=f^{\dagger}_{\alpha,j}f_{\alpha,j}$ ($\alpha = $A, B, C, D).
%\csouttmthd{In this case, $|\Psi_{\rm L}\rangle$ is not necessarily the ground state but is embedded at non-special position in the many-body spectrum. }
Hereafter we set $t_1=1$.
%%%%%%%%%%%%%%%%%%%%%%%%%%%%%%%%%%%%
%Fig
%\widetext
\begin{figure}[h]
\centering
%\begin{center}
%\centering  
\includegraphics[width=0.95\linewidth]{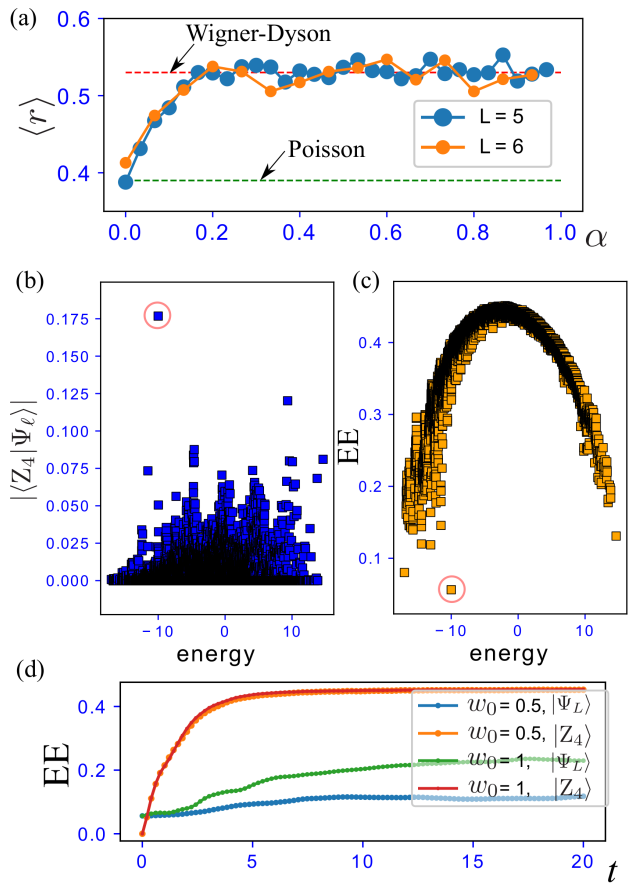}
%\includegraphics[width=7.5cm]{Fig2.eps}
%\end{center} 
% \hspace{10truemm}
\caption{Numerical results for the saw-tooth lattice model. 
(a) Mean level spacing $\langle r\rangle$ averaged over all energy eigenvalues in the momentum sector $k=0$ (See \cite{Santos,Turner}). 
We set $L=5$ with five particles and $L=6$ with six particles. 
(b) Overlap to $|{\rm Z}_4\rangle$ for all eigenstates. 
The red circle indicates $|\Psi_{\rm L}\rangle$. 
(c) Distribution of EE. The EE is normalized by the number of sites in the subsystem. The red circle indicates $|\Psi_{\rm L}\rangle$.
The EE for the red circle is very small because the only single CLS is cut, the value is 
$s_{\rm min}=\frac{1}{2L}(2\ln 2-\frac{3}{4}\ln3) \approx 0.05623$. 
For (b) and (c), we set $\mu=-2$, $V_0=3$, and $L=5$ with five particles. (d) Single-shot quench dynamics of EE: the disorder is $\mu_{A,j}$, $\mu_{B,j} \in [-w_0,w_0]$ with $V_0=3$. 
We set $L=5$ with five particles and two initial states, $|\Psi_{\rm L}\rangle$ and $|{\rm Z_4}\rangle$.
}
\label{Fig2}
\end{figure}
%%%%%%%%%%%%%%%%%%%%%%%%%%%%%%%%%%%%

As a first step, we demonstrate the nonintegrability from level spacing analysis. To be concrete, we calculate the level spacing $r_s$ defined by $r_{s}=[{\rm min}(\delta^{(s)}, \delta^{(s+1)})]/[{\rm max}(\delta^{(s)},\delta^{(s+1)})]$ for all $s$, where $\delta^{(s)}=E_{s+1}-E_{s}$ and $\{E_{s}\}$ is the set of energy eigenvalue (in ascending order) and calculate the mean level spacing $\langle r\rangle$ which obtained by averaging over $r_{s}$ by employing all energy eigenvalues with fixed momentum space.
By introducing a parameter $\alpha$ as $\mu=-2\alpha$ and $V_0=3\alpha$, we observe the change of the integrability.  
Figure \ref{Fig2}(a) is the numerical result. 
As increasing $\alpha$, $\langle r\rangle$ shows 
a clear crossover from integrable ($\langle r\rangle\simeq 0.39$, corresponding to the Poisson distribution) to nonintegrable ($\langle r\rangle\simeq 0.53$, corresponding to the Wigner-Dyson distribution). 
Hence, the term $H_{\rm int}$ makes the system nonintegrable. 
In what follows, we focus on the nonintegrable parameter point $\alpha=1$. 

To verify the presence of the QS state $|\Psi_{\rm L}\rangle$, 
we next calculate an overlap $|\langle {\rm Z}_4| \Psi_{\ell}\rangle|$, 
where $|{\rm Z}_4\rangle=\prod_{j=0}f^{\dagger}_{C,j}|0\rangle$ and $|\Psi_\ell\rangle$ is a many--body eigenstate for $H_{\rm tot}$.
As a typical character, we expect that $|\Psi_{\rm L}\rangle$ has large overlap compared to the other eigenstates.
The result is shown in Fig.~\ref{Fig2}(b). 
We find an atypical state with a large overlap, which is nothing but $|\Psi_{\rm L}\rangle$.

Next, we divide the system into two parts where both parts include $2L$ lattice sites and calculate the EEs of all eigenstates for the subsystem \cite{EE_cut}. 
The result is shown in Fig.~\ref{Fig2}(c). 
We find that the QS state $|\Psi_{\rm L}\rangle$ embedded in the energy excitation band 
exhibits very low-valued EE while other eigenstates have large value of the EE and show an arched distribution, 
which is typical a character of thermalized states in various system \cite{Ok,Wildeboar,Srivatsa}. Actually, the value of the EE for $|\Psi_{\rm L}\rangle$ can be easily obtained from cutting the single CLS. The simple calculation is shown in the Supplemental Material~\cite{Supp}. The numerical result of the EE for the QS agrees with the analytical result. 

Additionally, we investigate effects of quench disorders for the system. 
Through observing EE in the quench dynamics of the system, we investigate: 
(I) How robust is the QS for clean system $|\Psi_{\rm L}\rangle$ to disorders? (II) Whether or not does a weakly disordered system have a similar QS state in the clean limit? 
In dynamics, we set two initial states $|\Psi_{\rm L}\rangle$ and 
$|{\rm Z_4}\rangle$ and calculate the unitary dynamics by using exact diagonalization. 
Here we introduce a random on-site potential, i.e., adding the following term:
$H_{\rm rand} = \sum_{j,\alpha} \mu_{j}^\alpha n_{j}^\alpha$ with $\mu_{j}^\alpha \in [-w_0,w_0]$.
We fix $\mu=0$ and $V_0=3$. 
In the presence of $H_{\rm rand}$, $|\Psi_{\rm L}\rangle$ is no longer an exact eigenstate. 
The result is presented in Fig.~\ref{Fig2}(d), where we see the EE of the initial $|{\rm Z_4}\rangle$ suddenly increases and reaches a saturation value. 
On the other hand, the growth of EE for the initial $|\Psi_{\rm L}\rangle$ is very slow. 
This implies that for the disordered system, there exists a QS which is very close to $|\Psi_{\rm L}\rangle$. We also investigate another type of disorder that keeps $|\Psi_{\rm L}\rangle$ an exact eigenstate; see the Supplemental Material \cite{Supp}.

{\it Extension to higher dimensions.---}
%------------------------------------------------------------------%
\begin{figure}[t]
\begin{center}
\includegraphics[width = 0.98\linewidth]{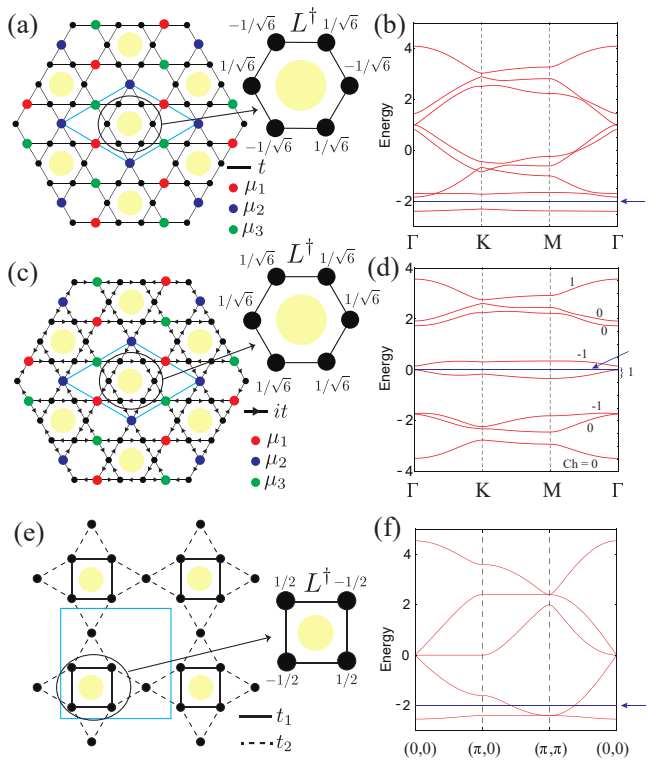}
\caption{(a) Schematic figure of the kagome model with real hoppings. 
The CLSs are located at the hexagons on which the yellow shaded balls are placed.
(b) The band structure for $(t, \mu_1, \mu_2, \mu_3) = (1, 0.5, 0.75,-0.8)$.
The high-symmetry points in the Brillouin zone are $\Gamma = (0,0)$, K$=\left( \frac{2\pi}{\sqrt{3}a_0},\frac{2\pi}{3a_0} \right)$,
and M$=\left( \frac{\pi}{\sqrt{3}a_0},\frac{\pi}{a_0} \right)$,
with $a_0$ being the lattice constant.
The blue line represents the flat band. 
(c) Schematic figure of the kagome model with pure imaginary hoppings.
The CLSs are located at the hexagons on which the yellow shaded balls are placed.
(d) The band structure for $(t,\mu_1, \mu_2, \mu_3) = (1,0.5, 0.75,-0.8)$. The blue line represents the flat band.
The Chern number is shown beside each band. 
(e) Schematic figure of the square kagome model.
The CLSs are located at the square plaquettes on which the yellow shaded balls are placed. 
(f) The band structure for $(t_1, t_2) = (1,1.2)$.
The blue line represents the flat band.}
 \label{fig:kagome_scar}
 \end{center}
\end{figure}
%------------------------------------------------------------------%
The construction of the QS is applicable in higher dimensions.
Here we present three concrete examples in two dimensions.

The first example is the real hopping model on a kagome lattice [Fig.~\ref{fig:kagome_scar}(a)].
Here we set the nearest-neighbor hopping $t$, being real, and the on-site potentials are set as  
$\mu_1$ for red dots, $\mu_2$ for blue dots, $\mu_3$ for green dots, and 0 otherwise.  
Due to the modulation of the on-site potential, the unit cell is enlarged compared with the conventional kagome model, resulting in nine sublattice degrees of freedom.
In this model, the CLSs reside on hexagons denoted by yellow shades in Fig.~\ref{fig:kagome_scar}(a),
whose wave function has a staggered sign structure~\cite{Zhitomirsky2004,Huber,Bilitewski2018,Rhim2019}.
Clearly, $d_m$ is equal to two, and thus the CLSs do not overlap each other.
This is a sharp contrast to the kagome model without on-site potentials, where CLSs live on all the hexagons and thus they overlap each other. 
The band structure is depicted in Fig.~\ref{fig:kagome_scar}(b).
We see an isolated flat band at $E = -2t$. 
Thus, when including the nearest-neighbor interaction and considering the $1/9$-filled system,
we obtained the QS in the form of Eq.~(\ref{eq:scarstate}).

The second example is the pure imaginary hopping model on a kagome lattice [Fig.~\ref{fig:kagome_scar}(c)]. 
The model without the on-site potential is investigated in the context of topological phase~\cite{Ohgushi2000,Else2019}.
In fact, the CLSs again reside on the hexagons but the wave function has a uniform sign structure. Then, we again find a flat band, as shown in Fig.~\ref{fig:kagome_scar}(d).
Moreover, the dispersive bands possess finite Chern number, 
which may lead to additional intriguing physics due to topology. 
Due to the CLSs, we can again construct the scar state at $1/9$-filling. 
Note that the flat band touches the dispersive band at the $\Gamma$ point, 
resulting in additional degeneracy for the non-interacting case.
Nevertheless, this additional degeneracy will be lifted when introducing the interaction, as the additional state is extended thus the many-body states occupying this state cost interaction energy. 

The third example is the square kagome model [Fig.~\ref{fig:kagome_scar}(e)].
The model has two hopping parameters, $t_1$ and $t_2$.
Remarkably, in this model, the CLSs with $d_m =2$ appear without incorporating the on-site potential.
To be concrete, the CLSs are on the square plaquettes, and have a staggered sign structure. 
The band structure is shown in Fig.~\ref{fig:kagome_scar}(f). 
We find a flat band corresponding to the CLSs, indicating the existence of the QS at $1/6$-filling.
Again, the flat band is degenerated with the dispersive band but it does not affect the emergence of the QS.
It would also be interesting to remark that the QS is found in the localized spin model on the square kagome lattice as well~\cite{McClarty}.

{\it Multiple QS.---}
Although we have discussed a unique QS, one can also construct a multiple QSs by reducing the number of the CLSs in Eq.~(\ref{eq:scarstate}). 
Also the saw-tooth lattice and the kagome lattice {\it without} on-site potentials. 

We further remark that our construction is extensible to systems with multiple flat bands (i.e., with different types of CLS), if we can fill CLSs spatially separated from each other. In such a case, possible patterns of filling CLSs become abundant, which results in multiple QSs. 
Candidates for such systems include metal organic frameworks (MOFs)~\cite{Kambe2014,Yamada2016,Kumar2018} and covalent organic frameworks (COFs) \cite{Shuku2018,Fujii2018,Cof}, where the tuning of electron filling may be feasible by using gate tuning or chemical doping.

{\it Conclusion.---}
We propose a general construction scheme of flat band QSs, making use of the ortho-normalized CLSs. 
As a simple example, we numerically demonstrate the presence of the unique QS for a saw--tooth lattice model, 
which can be implemented in coldatom optical lattice systems \cite{Saw-tooth_lattice1,Saw-tooth_lattice2}.
We also present some examples of higher dimensions, namely, kagome and square kagome lattice systems. 
We expect that our construction of the unique QS for non-overlapping CLSs has very wide-range applications for a Wigner-crystal state on a $p_{x-y}$-orbital honeycomb lattice system \cite{Wu}, and not only for fermions but also for hard-core bosons, which is related to the recent experiments on Rydberg atoms~\cite{Leseleuc}. Further, as we have mentioned above, searching QSs in chemical systems such as MOFs and COFs will be another interesting direction. 
Thus, our work will open up a new way for realizing QSs in broader class of systems. 

This work is supported in part by JSPS
KAKENHI Grant Numbers JP17H06138 (Y.K, Y.H.) and JP20K14371 (T.M.).
%%%%%%%%%%%%%%%%%%%%%%%%%%%%%%%%%%%%%%%%%%%%%%%

\clearpage

%%%%%%%%%%%%%%%%%%%%%%%%%%%%
% Supplemental material
%%%%%%%%%%%%%%%%%%%%%%%%%%%%
\section*{Supplemental material}
\subsection{Comparison with Shiraishi-Mori construction}
We show a relationship to a construction method for the scar state by Shiraishi and Mori \cite{Shiraishi}.
They considered a nonintegrable system (Eq.~(2) in \cite{Shiraishi}):
\begin{eqnarray}
H_{\rm SM}= \sum_{j}P_j h_{j}P_j + H'.  \nonumber
\end{eqnarray}
Here, according to \cite{Shiraishi}, $P_{j}$ is arbitrary local projection operators and a scar state $|\Psi_s\rangle$ can be introduced by $P_{j}|\Psi_{s}\rangle=0$.
Further, $H'$ is a Hamiltonian which satisfies $[P_j,H']=0$.
Actually, if we consider the Hamiltonian  
$\tilde{H} = H_{\rm tot} - \varepsilon_{\rm L} \sum_{j} L^\dagger_j L_j$ 
with $\varepsilon_{\rm L}$ being the flat band energy, 
we find a correspondence for our construction in a flat--band system: 
\begin{eqnarray}
1- L^{\dagger}_j L_j&\Longleftrightarrow& P_j, \\
\tilde{H}  &\Longleftrightarrow& H', \\
-\varepsilon_{\rm L} \hat{I}_{j} &\Longleftrightarrow& h_j,\\
|\Psi_{\rm L}\rangle &\Longleftrightarrow& |\Psi_s\rangle,
\end{eqnarray} 
Certainly, $[1-L^{\dagger}_{j}L_{j}]|\Psi_{L}\rangle=0$, 
but our construction is slightly relaxed in that the relation $[P_j,H']=0$ is not satisfied as an operator relation, that is, in our scar state $|\Psi_{\rm L}\rangle$ is not necessarily vanishing when operating $\tilde{H}$
but it can be an eigenstate of $\tilde{H}$ with an arbitrary eigenvalue.
Accordingly, our scar construction for a flat band system is a little different from the Shiraishi-Mori construction in that 
the operator relation $[P_j,H']=0$ is not required, but the SQ is reqired to be an eigenstate for $H'$. Similar comparison has been discussed in some topological models \cite{Ok}.\\ 

\subsection{Single particle spectrum for saw--tooth lattice}
\begin{figure}[h]
\centering
%\begin{center}
%\centering  
\includegraphics[width=4cm]{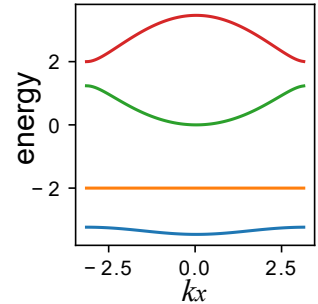}
%\end{center} 
% \hspace{10truemm}
\caption{Single particle energy spectrum for the saw--tooth lattice with an on-site potential $\mu=-2$. Here, we set the momentum $k_x \in [-\pi,\pi)$, where the lattice spacing is unity.} 
\label{FigS1}
\end{figure}

We show the single particle spectrum for the saw-tooth-lattice model. The non--interacting Hamiltonian is  
\begin{eqnarray}
H_0&=&\sum^{L-1}_{j=0} \biggl[t_1 f^{\dagger}_{j,\mathrm{A}} f_{j,\mathrm{C}}+t_2f^{\dagger}_{j,\mathrm{A}}f_{j,\mathrm{B}}\nonumber\\
&+&t_2f^{\dagger}_{j,\mathrm{B}}f_{j,\mathrm{C}}
+t_1 f^{\dagger}_{j,\mathrm{C}} f_{j+1,\mathrm{A}}\nonumber\\
&+&t_2f^{\dagger}_{j,\mathrm{C}}f_{j,\mathrm{D}}
+t_2f^{\dagger}_{j,\mathrm{D}}f_{j+1,\mathrm{A}}
+\mbox{h.c.}\biggl] \nonumber\\
&+& \sum^{L-1}_{j=0}\mu f^{\dagger}_{j,\mathrm{A}}f_{j,\mathrm{A}}.
\label{HST}
\end{eqnarray}
We set $t_1=1$,$t_2=\sqrt{2}$ and $\mu=-2$. 
See Fig.~\ref{FigS1}, the spectrum from the bulk momentum representation exhibits four bands where the second lowest band is flat and the others are dispersive. 
Also by varying the sign and/or value of the hopping and $\mu$, one can control the position of the flat band in the spectrum.

%%%%%%%%%%%%%%%%%%%%%%%%%%%%%%%%%%%%
%Fig
%\widetext
%%%%%%%%%%%%%%%%%%%%%%%%%%%%%%%%%%%%

\subsection{Molecular orbital picture for saw--tooth lattice model}
With the help of the molecular orbital(MO) picture, 
we verify the existence of the single isolated flat band for the saw-tooth-lattice model discussed in the main text. 

Here, we introduce the following three MOs, 
\begin{eqnarray}
C^{\dagger}_{1,j}&=&f^{\dagger}_{\mathrm{A},j}+\sqrt{2}f^{\dagger}_{\mathrm{B},j}+f^{\dagger}_{\mathrm{C},j},\\
C^{\dagger}_{2,j}&=&f^{\dagger}_{\mathrm{C},j}+\sqrt{2}f^{\dagger}_{\mathrm{D},j}+f^{\dagger}_{\mathrm{A},j+1},\\
C^{\dagger}_{3,j}&=&f^{\dagger}_{A,j}.
\end{eqnarray}
Note that $C_{1,j}$, $C_{2,j}$, and $C_{3,j}$ are linearly independent of each other.
We also note that all of these MOs anticommute the CLS: $\{ C_{\eta,j} , L^\dagger_{j^\prime}\} = 0$ with $\eta  =1,2,3$.
Then, by using these orbitals, the Hamiltonian of Eq.~(\ref{HST}) can be rewritten as
\begin{eqnarray}
H_{0}&=&\sum^{L-1}_{j=0}\biggl[t_1C^{\dagger}_{1,R_j}C_{1,R_j}
+t_1C^{\dagger}_{2,R_j}C_{2,R_j}\nonumber\\
&&+\mu C^{\dagger}_{3,R_j}C_{3,R_j}\biggr] - 2t_1\sum_{j,\alpha} n_{j}^{\alpha}. \label{eq:ST_MOrep}
\end{eqnarray}
Note that the last term of Eq.~(\ref{eq:ST_MOrep}) gives an mere entire shift of single-particle energy. 
According to Refs.~\cite{Maruyama,Mizoguchi}, 
the number of the linearly-independent molecular orbital states infers the number of zero energy mode in the flat band system. 
Namely, since the number of atomic orbitals is $4L$ whereas that of the MOs is $3L$, the remaining $L$ degrees of freedom 
become zero modes of $H_{0} - 2t_1\sum_{j,\alpha} n_{j}^{\alpha}$, i.e., the flat band with energy $-2t_1$ of $H_{0}$.
In fact, these remaining degrees of freedom are described by the CLSs.

%%%%%%%%%%%%%%%%%%%%%%%%%%%%%%%%%%%%
%Fig
%\widetext
\begin{figure}[t]
\centering
%\begin{center}
%\centering  
\includegraphics[width=8cm]{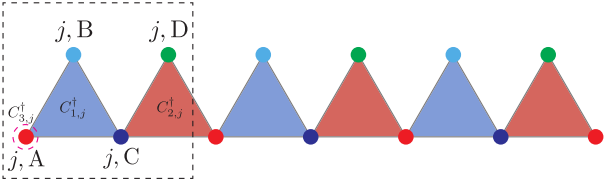}
%\end{center} 
% \hspace{10truemm}
\caption{Molecular orbitals in saw-tooth lattice.
The red and blue triangular objects and the red dotted circle are $C^{\dagger}_{1,j}$, $C^{\dagger}_{2,j}$ and $C^{\dagger}_{3,j}$.}
\label{FigS2}
\end{figure}
%%%%%%%%%%%%%%%%%%%%%%%%%%%%%%%%%%%%

\subsection{Entanglement entropy for a single CLS in saw--tooth lattice}
\begin{figure}[b]
\centering
%\begin{center}
%\centering  
\includegraphics[width=2.2cm]{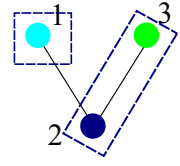}
%\end{center} 
% \hspace{10truemm}
\caption{Schematic figure for the entanglement cut for a single CLS. 
The blue dotted boxes represent subsystems.}
\label{FigS3}
\end{figure}

%(This is a note, so this is not necessary in the supplemental material...)
We can calculate the EE for $|\Psi_{\rm L}\rangle$. 
Here, we assume $L= odd$ case. 
To calculate the EE, we divide the system into two halves as mentioned in the main text. 
At the time of the division, only single CLS is cut.
Only this entanglement cut contributes to the EE of $|\Psi_{\rm L}\rangle$. 
Accordingly, we focus on calculating the EE of a single CLS on three site labeled by 1,2,3 as shown in Fig.~\ref{FigS3}.
Then, the density matrix for the single CLS is given by
\begin{eqnarray}
\rho_{C}&=&L_{j=(L-1)/2}^{\dagger}|0\rangle \langle 0|L_{j=(L-1)/2}\nonumber\\
=&&\frac{1}{2}|010\rangle\langle 010|
-\frac{\sqrt{2}}{4}|010\rangle\langle 100|
-\frac{\sqrt{2}}{4}|010\rangle\langle 001|\nonumber\\
&&-\frac{\sqrt{2}}{4}|100\rangle\langle 010|
-\frac{1}{4}|100\rangle\langle 100|
-\frac{1}{4}|100\rangle\langle 001|\nonumber\\
&&-\frac{\sqrt{2}}{4}|010\rangle\langle 001|
-\frac{1}{4}|100\rangle\langle 001|
-\frac{1}{4}|001\rangle\langle 001|,\nonumber\\
\end{eqnarray}
where $|abc\rangle$ ($a+b+c=1$) is a particle number state for three site system, e.g., $|010\rangle$ means that the particle resides at the site 2. 
Here, we set a subsystem including only the site 1, and calculate the EE for the subsystem.
By taking the trace for the particle states for the sites $2$ and $3$, the reduced density matrix is 
\begin{eqnarray}
{\rm Tr}_{(2,3)}\rho_{C}=\frac{3}{4}|1\rangle_{1}\langle 1|+\frac{1}{4}|0\rangle_{1}\langle 0|.
\end{eqnarray} 
Accordingly, the EE for this entanglement cut is given by $s_{\rm CLS}=2\ln 2-\frac{3}{4}\ln 3$, 
corresponding to the EE of $|\Psi_{\rm L}\rangle$.
This value agrees with the numerical result in Fig.~\ref{Fig2} (c) in the main text. 

%%%%%%%%%%%%%%%%%%%%%%%%%%%%%%%%%%%%
%Fig
%\widetext
%%%%%%%%%%%%%%%%%%%%%%%%%%%%%%%%%%%%

\subsection{Disorder effect to dynamics of entanglement entropy in saw--tooth lattice} 
%%%%%%%%%%%%%%%%%%%%%%%%%%%%%%%%%%%%
%Fig
%\widetext
\begin{figure}[b]
\centering
%\begin{center}
%\centering  
\includegraphics[width=0.95\linewidth]{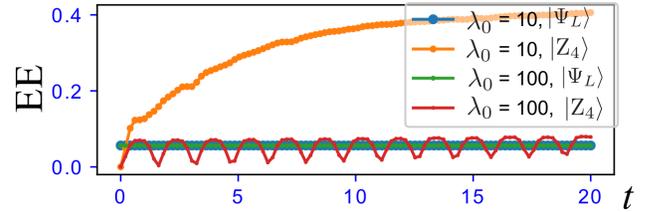}
%\includegraphics[width=8cm]{FigS4.eps}
%\end{center} 
% \hspace{10truemm}
\caption{Single-shot dynamics of EE: The case of the quench on-site disorder $\mu_j \in [-\lambda_0,\lambda_0]$ and interaction disorder $V_{0}\to V_j\in [-\lambda_0,\lambda_0]$. 
We set $L=5$ with five particles. 
We set two initial states, $|\Psi_{\rm L}\rangle$ and $|Z_4\rangle$. We set $t_1=1$, $\mu=-2$.}
\label{FigS4}
\end{figure}
%%%%%%%%%%%%%%%%%%%%%%%%%%%%%%%%%%%%
We show effects of quench disorders that keep $|\Psi_{\rm L}\rangle$ an exact eigenstate. 
To be concrete, for the system of Eq.~(\ref{HST}) with the interaction of Eq.~(\ref{int_ST}) in the main text, we introduce a disorders in the following manner:
$\mu \to \mu_j$, and $V_{0}\to V_j$ where $\mu_j$, $V_j \in [-\lambda_0,\lambda_0]$ (uniform disorder). 
In dynamics, we set two initial states $|\Psi_{\rm L}\rangle$ and $|{\rm Z_4}\rangle$. 
Figure~\ref{FigS4} is the dynamics of EE for the subsystem defined in the main text. While the EE for the initial $|{\rm Z_4}\rangle$ state increases during time evolution although the large disorder suppresses the increase, the EE for the initial $|\Psi_{\rm L}\rangle$ does not change at all. 
This indicates that $|\Psi_{\rm L}\rangle$ is indeed still the QS even for the disordered system.


\begin{thebibliography}{99}
\bibitem{gene_ham}
Here, this model is for any spatial dimension and lattice structure.
\bibitem{Uni}
Here, the state $|\Psi_{\rm L}\rangle$ is unique since it is created by the entire occupation of the flat band. 
The degenerate scar state will be discussed in future work.
\bibitem{EE_cut}
For example, if we set $L=5$ system, the subsystem includes the lattice site, $(j,\alpha)=\{(0,A),(0,B),(0,C),(0,D),\cdots, ((L-1)/2,A),((L-1)/2,B)\}$.
\end{thebibliography}
\end{document}